\documentclass{raa}            % referee version: for submission

\usepackage{graphicx,times}       
\usepackage{natbib}
\usepackage{amssymb,amsmath}
\bibpunct{(}{)}{;}{a}{}{,}
\usepackage[pagebackref=true]{hyperref}

\begin{document}

  \title{Polarization Studies in Different Blazar Types. Part I: Correlation between Optical Brightness and Polarization Degree
  }

   \volnopage{Vol.0 (20xx) No.0, 000--000}      %% preserved for Editor. DOn't remove!
   \setcounter{page}{1}          %% starting page, preserved for Editor. DOn't remove!

   \author{E. A. Shkodkina
      \inst{1}
   \and S. S. Savchenko
      \inst{1, 2}
  \and D.~A.~Morozova\inst{1}
  \and S.~G.~Jorstad\inst{1, 3}
  \and G.~A.~Borman\inst{4}
  \and T.~S.~Grishina\inst{1}
  \and E.~N.~Kopatskaya\inst{1}
  \and E.~G.~Larionova\inst{1}
  \and A.~A.~Vasilyev\inst{1}
  \and I.~S.~Troitskiy\inst{1}
  \and Yu.~V.~Troitskaya\inst{1}
  \and P.~A.~Novikova\inst{1}
  \and E.~V.~Shishkina\inst{1}
  \and A.~V.~Zhovtan\inst{4}
   }

   \institute{Saint Petersburg State University, 7/9 Universitetskaya nab., St. Petersburg, 199034 Russia.;
   {\it shkodkina.e.a@gmail.com} \\
%% Please give the E-mail address of the author, to whom future correspondence andboffprint requests will be sent.
    \and
         Central (Pulkovo) Astronomical Observatory, Russian Academy of Sciences,
         St. Petersburg, 196140, Russia \\
    \and
        Institute for Astrophysical Research, Boston University, 725 Commonwealth Avenue, Boston, MA 02215, USA \\
    \and
        Crimean Astrophysical Observatory RAS, P/O Nauchny, 298409, Russia \\
\vs\no
   {\small Received 20xx month day; accepted 20xx month day}}

\abstract{We present an analysis of the cross-correlation between optical brightness and polarization degree in different types of blazars. The aim is to identify objects with simultaneous and consistent changes in characteristics and to determine if this behavior relates to the types of objects studied. The analysis includes 23 objects: 11~FSRQ, 11~BL~Lac, and 1 radio galaxy. Dense overlapping observation series in the \textit{R} band were used, collected over more than 10~years as part of a monitoring program for bright blazars at St.~Petersburg State University. The cross-correlation analysis procedure is detailed, including a method for assessing significance based on Monte Carlo simulations of synthetic light curves modeled using a Damped Random Walk. Significant correlations were found for 5~FSRQ and 1~BL~Lac. No significant correlation was detected for 10~BL~Lac and 6~FSRQ. One object did not yield a reliable estimate. Based on the current results, we cannot claim that the observed difference in the behavior of these emission characteristics for different classes of blazars is significant. It is possible that observed correlations may be explained by the contribution of simultaneous flare events to the changes in flux and polarization degree curves, which occur more frequently in FSRQ objects.
\keywords{galaxies: active --- BL Lacertae objects: general --- polarization --- methods: data analysis
}
}

   \authorrunning{E. A. Shkodkina et al.}   
   %author_head in even pages
   \titlerunning{Optical Brightness \& Polarization Correlation}  
   % title_head in odd pages

   \maketitle

\section{Introduction}   
\label{sect:intro}
Blazars are a class of Active Galactic Nuclei (AGN) with extremely bright, variable emission, dominating over the host galaxy. Variability occurs on timescales from minutes to decades~\citep{hufnagel_1992}. Unified models suggest this emission originates from a relativistic jet pointed at a small angle to the observer, resulting in relativistic beaming~\citep{blandford_1978}. Blazars are divided into FSRQs (Flat Spectrum Radio Quasars) and BL Lac objects, which differ in their optical spectra~\citep{urry_1995}. FSRQs show strong, broad emission lines, while BL Lac objects have weak or no lines.

Blazar spectral energy distributions feature two components: a low-energy one peaking in the optical, ultraviolet, or X-ray range, and a high-energy one peaking in the gamma range. Both are considered non-thermal. The low-frequency component is attributed to synchrotron radiation from relativistic electrons in the jet's magnetic field~\citep{marscher_2008}. It is believed that the synchrotron nature causes high polarization, with degrees up to $50\%$~\citep{mead_1990}. However, many details of the variable polarized emission generation mechanisms in blazars remain unclear.

A variety of jet models have been proposed to explain the observed blazar emission variability. These include models based on chaotic magnetic fields compressed by shock waves~\citep{laing_1980, angelakis_2016}, helical jets with helical magnetic fields~\citep{raiteri_2013}, somehow related to the spin of the central black hole~\citep{chen_2024}, and models involving knots moving in a toroidal magnetic field~\citep{marscher_2010}. Other models consider magnetic field line reconnection, leading to energy release and particle acceleration on short time-scales~\citep{sironi_2014}. A model for large-scale flux variations involves changes in Doppler boosting due to temporal variations in the jet viewing angle~\citep{2017Natur.552..374R}. Models based on turbulent emitting cells in jets are also used to explain the chaotic variability of polarized emission~\citep{marscher_2014}.

Studying the variability of polarized emission is an accessible way to obtain information about the magnetic field, its role in particle acceleration, its interaction with the surrounding environment, and its relation to overall emission. Examining the connection between events with high and low polarization can provide understanding of the central regions' structure and the processes occurring within them.

Many studies have focused on analyzing the relationship between the degree of polarization and total flux. For example, in~\cite{raiteri_2012, itoh_2018, otero_2023} correlated evolution of the polarization degree and the total flux was reported for different objects. In other cases, an anticorrelation was observed~\citep{fraija_2017, pandey_2021, rajput_2022, bachev_2023}. Also no relationship was found~\citep{covino_2015, itoh_2018, otero_2023}. The data and processing methods used in each of these studies differed significantly, which may contribute to the variability in the reported findings.  

This work focuses on studying the correlation between variations in the polarization degree and brightness in the optical spectrum for various types of blazars. We analyze long observation series, with particular emphasis on the methods used for cross-correlation analysis. The structure of the paper is as follows: Section~\ref{sect:obs} provides a description of the sample and observation process. Section~\ref{sect:methods} explains the methodology applied for data analysis. The main results and discussion are presented in Sections~\ref{sect:results} and~\ref{sect:discussion}, respectively.

\section{Sample and Observations}
\label{sect:obs}

In this work, we study gamma-ray bright blazars included in the observational program at Saint Petersburg State University\footnote{\url{https://vo.astro.spbu.ru/program_all/}}. We possess extended observational series compiled from data obtained by various telescopes, as detailed below. Observations spanning 10 years or more provide unique datasets with long durations and dense measurements of polarization characteristics for most of the objects included.

A total of 23~objects were considered in the study: 11~FSRQ-type, 11~BL~Lac-type and 1~radio galaxy (RG). For these objects, there are overlapping optical observations of polarization characteristics and total flux. Types of objects (according to those specified in the monitoring program and their main characteristics (according to NASA/IPAC Extragalactic Database\footnote{\url{http://ned.ipac.caltech.edu/}}) are listed in Table~\ref{tab:objects}. Information about the optical total and polarized flux observations are provided in Table~\ref{tab:series}.

\begin{table}[ht!]
\begin{center}
\caption[]{Information about the objects.}\label{tab:objects}
\begin{tabular}{lcccc}
\hline\noalign{\smallskip}
Name & Type & RA & DEC  & z \\
 &  & (J2000) & (J2000) &  \\
\hline\noalign{\smallskip}
3C 66a & BLLac & 02 22 39.6 & +43 02 08 & 0.370000 \\ 
OJ 049 & BLLac & 08 31 48.9 & +04 29 39 & 0.173856 \\ 
Mkn 421 & BLLac & 11 04 27.3 & +38 12 32 & 0.030021 \\ 
S4 0954+65 & BLLac & 09 58 47.2 & +65 33 55 & 0.369400 \\ 
BL Lacertae & BLLac & 22 02 43.3 & +42 16 40 & 0.066800 \\ 
OJ 287 & BLLac & 08 54 48.9 & +20 06 31 & 0.305600 \\ 
PG 1553+11 & BLLac & 15 55 43 & +11 11 24 & 0.360000 \\ 
PKS 0735+17 & BLLac & 07 38 07.4 & +17 42 19 & 0.424000 \\ 
Q 1959+65 & BLLac & 19 59 59.8 & +65 08 54 & 0.047000 \\ 
S5 0716+71 & BLLac & 07 21 53.4 & +71 20 36 & 0.310000 \\ 
W Com & BLLac & 12 21 31.7 & +28 13 59 & 0.102000 \\ 
AO 0235+16 & FSRQ & 02 38 38.9 & +16 36 59 & 0.940000 \\ 
3C 454.3 & FSRQ & 22 53 57.7 & +16 08 54 & 0.859000 \\ 
OJ 248 & FSRQ & 08 30 52.1 & +24 11 00 & 0.938770 \\ 
CTA 102 & FSRQ & 22 32 36.4 & +11 43 51 & 1.032000 \\ 
3C 273 & FSRQ & 12 29 06.7 & +02 03 09 & 0.158339 \\
3C 279 & FSRQ & 12 56 11.1 & -05 47 22 & 0.536200 \\ 
CTA 26 & FSRQ & 03 39 31 & -01 46 36 & 0.852000 \\ 
PKS 0420-01 & FSRQ & 04 23 15.8 & -01 20 33 & 0.916090 \\ 
Q 0836+71 & FSRQ& 08 41 24.3 & +70 53 42 & 2.172000 \\ 
Q 1156+29 & FSRQ & 11 59 31.8 & +29 14 44 & 0.724745 \\ 
PKS 1222+21 & FSRQ & 12 24 54.4 & +21 22 46 & 0.433826 \\ 
3C 84 & RG & 03 19 48.2 & +41 30 42 & 0.017559 \\ 
\noalign{\smallskip}\hline
\end{tabular}
\end{center}
\end{table}

\begin{table}[ht!]
\begin{center}
\caption[]{Information about the optical total and polarized flux observations.}\label{tab:series}
\begin{tabular}{lcccccccc}
\noalign{\smallskip}
\hline
Name & \multicolumn{3}{c}{R} & & \multicolumn{3}{c}{Polarization Degree} \\
\cline{2-4} \cline{6-8}
            & Num. points & Length & Mean Cadence & & Num. points & Length & Mean Cadence \\
            &  & (days) & (days) & & & (days) & (days) \\
\hline
\noalign{\smallskip}
3C 66a & 1902 & 6390.41 & 3.36 & & 2170 & 6390.41 & 2.95 \\ 
OJ 049 & 429 & 5793.17 & 13.54 & & 429 & 5793.17 & 13.54 \\ 
Mkn 421 & 1677 & 4712.49 & 2.81 & & 2121 & 5187.93 & 2.45 \\ 
S4 0954+65 & 2016 & 6590.11 & 3.27 & & 2028 & 6590.11 & 3.25 \\
BL Lacertae & 6200 & 6563.94 & 1.06 & & 6553 & 6563.94 & 1.00 \\
OJ 287 & 2569 & 6606.06 & 2.57 & & 2953 & 6606.06 & 2.24 \\ 
PG 1553+11 & 688 & 3213.14 & 4.68 & & 741 & 3213.14 & 4.34 \\
PKS 0735+17 & 671 & 6354.71 & 9.48 & & 710 & 6354.71 & 8.96 \\ 
Q 1959+65 & 716 & 5420.06 & 7.58 & & 844 & 5420.06 & 6.43 \\ 
S5 0716+71 & 4983 & 6568.12 & 1.32 & & 5204 & 6568.12 & 1.26 \\
W Com & 881 & 6606.0 & 7.51 & & 1143 & 6606.0 & 5.78 \\
AO 0235+16 & 989 & 6245.32 & 6.32 & & 1151 & 6245.32 & 5.43 \\ 
3C 454.3 & 2159 & 5870.36 & 2.72 & & 2622 & 5870.36 & 2.24 \\
OJ 248 & 383 & 5788.43 & 15.15 & & 521 & 5788.43 & 11.13 \\ 
CTA 102 & 2167 & 6410.1 & 2.96 & & 2310 & 6410.1 & 2.78 \\ 
3C 273 & 509 & 5847.43 & 11.51 & & 747 & 5847.43 & 7.84 \\
3C 279 & 1845 & 5852.0 & 3.17 & & 2188 & 5852.0 & 2.68 \\ 
CTA 26 & 409 & 5885.3 & 14.42 & & 431 & 5885.3 & 13.69 \\ 
PKS 0420-01 & 434 & 6546.48 & 15.12 & & 447 & 6546.48 & 14.68 \\
Q 0836+71 & 402 & 5800.51 & 14.47 & & 407 & 5800.51 & 14.29 \\
Q 1156+29 & 1482 & 6815.58 & 4.60 & & 1552 & 6815.58 & 4.39 \\ 
PKS 1222+21 & 803 & 5591.46 & 6.97 & & 1105 & 5591.46 & 5.06 \\ 
3C 84 & 301 & 3697.56 & 12.33 & & 301 & 3697.56 & 12.33 \\ 
\noalign{\smallskip}\hline
\end{tabular}
\end{center}
\end{table}

We combine optical photometric and polarimetric data obtained using following telescopes: LX-200 ($40\,$cm, Saint Petersburg State University, Peterhof), AZT-8 ($70\,$cm, Crimean Astrophysical Observatory, Nauchny), Perkins ($1.83\,$m, Perkins Telescope Observatory, Flagstaff, AZ, USA), and the Kuiper and Bok telescopes ($1.54\,$m and $2.29\,$m, respectively, Steward Observatory, Tucson, AZ, USA).

Measurements at the LX-200 telescope were carried out without a filter until late 2018, with a central wavelength of $\lambda_{\text{eff}}=6700\,$\AA; afterward, the \textit{R} filter was used. The instrumental polarization was measured using stars located near the object in the celestial sphere, assuming that the radiation from the calibration stars is unpolarized. Generally, the measurement errors in the polarization degree do not exceed~$1\%$ for objects with a magnitude of~$\sim17^m$. The typical total exposure times (summed over several individual exposures) range from $100\,$s for bright objects to $1500\,$s for faint ones. Median polarization degree for objects in our sample is $7.9\%$, with the values $15.2\%$ for the object 3C~279 with typically high polarization degree and $0.3\%$ for the 3C~273 with typically low polarization degree. 

The observations at Steward Observatory were performed using the CCD Imaging/Spectro-polarimeter (SPOL;~\cite{schmidt_1992}), yielding spectra from $4000\,$\AA \ to $7550\,$\AA. The spectra were used to calculate the linear polarization parameters within $5000 - 7000\,$\AA, close to \textit{R} band. The majority of spectra were accompanied by photometric measurements in the \textit{V} band~\citep{schmidt_1992}, which are not included in our study, as we aim to ensure data uniformity and focus on the analysis of brightness in the \textit{R} filter specifically.  Some Steward data provide the total flux density in \textit{R} band; \textit{V} photometry was used to normalize spectra and then from a spectrum the brightness in \textit{R} band. Observations at other instruments were conducted in the \textit{R} filter. 

Detailed information on the methodology for collecting and processing observations from the LX-200 and AZT-8 telescopes is available in~\cite{larionov_2008}, for the Perkins telescope in~\cite{jorstad_2010} and for the Steward observatory monitoring programm in~\cite{smith_2009}.

\section{Analysis Methods}
\label{sect:methods}
For the cross-correlation analysis, we use the Local Discrete Correlation Function (LDCF) --- a modification of the Discrete Correlation Function~\cite{edelson_1988} algorithm with local normalization as proposed by~\cite{welsh_1999}. Additionally, we provide an assessment of the significance of cross-correlation. A commonly used method is to calculate confidence intervals via Monte Carlo (MC) simulations of the light curves. In the current study, we propose using the Damped Random Walk (DRW) model for simulating the light curves.

\subsection{Cross-Correlation Estimation}
The Cross Correlation Function (CCF) quantifies the statistical relationship between two variables over time. It is commonly used to determine the similarity between two time series on different time shifts (also often referred as lags). CCF of two real stationary stochastic processes is defined as:
\begin{equation*}\label{eq:corr_real}
R_{XY}(\tau) = \frac{\mathbb{E}[(x(t)-\mu_{X})(y(t+\tau)-\mu_{Y})]}{\sigma_{X} \sigma_{Y}},
\end{equation*}
where $\mathbb{E}[.]$ is the expected value, $\tau$ is the time lag, and $x(t)$ and $y(t)$ are realizations of the processes $X(t)$ and $Y(t)$. The means and standard deviations are $\mu_{X}, \mu_{Y}$ and $\sigma_{X}, \sigma_{Y}$, respectively. The function is normalized so that $R_{XY}$ takes values between $[-1;1]$. 

In practical applications, finite discrete realizations of random processes are typically observed. To estimate sample cross-correlation in discrete cases, various approximations of this definition are used. The cross-correlation estimation for stationary time series $x(t)$ and $y(t)$, whose values are uniformly distributed over time as $t_n = (n - 1)\Delta t$, where $n = 1, 2, \ldots, N$, can be expressed as:
\begin{equation*}\label{eq:corr_ast}
    r_{xy}(\tau_k) = \frac{\frac{1}{N^*} \sum\limits_{n=1}^{N}(x_n - \bar{x})(y_{n+k} - \bar{y})}{\sqrt{\frac{1}{N} \sum\limits_{n=1}^{N} (x_n - \bar{x})^2 \frac{1}{N} \sum\limits_{n=1}^{N} (y_{n} - \bar{y})^2}}.
\end{equation*}
In this case, the lag takes only discrete values $\tau_k = k\Delta t$, where $k \in \mathbb{Z}$. Here, $\bar{x}$ and $\bar{y}$ are the sample means of $x(t)$ and $y(t)$. The coefficient $\frac{1}{N^*}$ may be equal $\frac{1}{N}$ or $\frac{1}{N-k}$, depending on chosen approximation~\citep{welsh_1999}. Although such estimation is correct, it is often impractical for real time series, which may be non-stationary, have measurement errors and irregular discretization. Additionally, measurement errors can vary due to changing conditions, emphasizing the need for methods that provide reliable estimates under these circumstances.

Various methods exist for estimating the correlation of uneven time series. Some researchers also estimate the cross-correlation of simultaneous series using the Spearman rank correlation coefficient. More universal mothods can be grouped into several categories: data interpolation, temporal data binnig, and Fourier transforms. All based on the assumption of series stationarity. In reality, assumptions about stationarity are almost always violated during extended observations of many real processes. This is particularly true for objects such as blazars, which can transition between active and quiescent states. Nevertheless, in the absence of alternative tools, researchers are often forced to overlook this assumption.

We performed a stationarity check on our data using the simple Dickey-Fuller test. For the $R$ band light curves of the objects 3C~273, 3C~84, CTA~102, Mkn~421, OJ~248, PKS 0735+17, and Q~1959+65, we failed to confirm stationarity. It is important to note that this test is not suitable for assessing the stationarity of uneven time series, so these results must be treated with particular caution. However, we lack alternative methods for analyzing uneven time series that do not require interpolation or more complex data transformations. We believe that we can still perform cross-correlation procedure, but it is important to approach the interpretation of the results with caution.

Our data consist of unevenly sampled time series with gaps and measurement uncertainties. For some objects, we have dense, accurate series, while for others, the observations are irregular and error-prone. In some cases, we also lack of photometric observations in \textit{R} band from the Steward monitoring program. Therefore, we need a cross-correlation estimation method that provides reliable results in such case. As noted by \citet{peterson_1993}, interpolation-based methods may fail with highly irregular sampling, and the commonly used ICCF method also does not account for measurement errors \citep{peterson_1998}. Fourier-based methods require data manipulation and mathematically complicate transitions. 

In our view, a reasonable choice in the current situation is the LDCF algorithm, which accounts for measurement errors and provides a more reliable, though conservative, estimate \citep{peterson_1993}. This method, along with the assessment of the significance algorithm used in this work, also avoids the need for transitioning to the frequency domain.

\subsubsection{Local Discrete Correlation Function}
In general, two data sets of different sizes are considered: time-ordered sequences of triples $X=\left\{t_{xi}, x_i, e_{xi}\right\}_{i=1}^N$ and $Y=\left\{t_{yj}, y_j, e_{yj}\right\}_{j=1}^K$. Each triple includes the observation time, the measured value, and its uncertainty. For every time lag $\tau$ an estimate of CCF for $X$ and $Y$ can be calculated:
\begin{equation*}\label{ldcf}
\text{LDCF}(\tau) = \frac{1}{M} \sum \frac{(x_i - \bar{x}_\tau)(y_j - \bar{y}_\tau)}{\sqrt{|(\sigma^2_{x\tau} - e^2_{xi})(\sigma^2_{y\tau} - e^2_{yj})|}}.
\end{equation*}

Lag $\tau$ lies in the center of a time interval $\Delta t$ with an arbitrary width. The sum is taken over \(M\) pairs of indices \((i, j)\) for which $\tau - \frac{\Delta t}{2} < \Delta t_{ij} < \tau + \frac{\Delta t}{2}$. The average values \(\bar{z}_\tau\) and standard deviations \(\sigma_{z\tau}\) are calculated only for the points within the corresponding time interval $\Delta t$. The resulting cross-correlation function values are limited to the interval [-1, 1].

The choice of time interval width $\Delta t$ represents a compromise between achieving an accurate estimate of the cross-correlation function and maintaining its detail. To obtain a reliable estimate, the number of points in each time interval must be large enough, general estimation is $M \gg 2$~\citep{edelson_1988}. 

The strength of the method lies in the fact that it requires no additional description or prediction of the time series behavior, using only the values and measurement errors without interpolation or complex transformations. A relative drawback is that numerous data pairs are required in each $\Delta t$ interval~\citep{peterson_1993}. If the sampling of the series under study differs significantly, choosing an appropriate interval width may become challenging. We suggest that these difficulties can be addressed by using an adaptive selection of time intervals, where the fixed parameter would not be the interval width $\Delta t$ but rather the required number of point pairs $M$. In the future, we hope to implement this method and compare the results of such an analysis with those presented here, obtained using fixed time intervals' width.

In this study, we were choosing fixed interval width $\Delta t$ for each object specifically, according to the sampling characteristic of its observations. In each case $M \gg 100$.

\subsection{Significance Assessment}
In the case of AGN light curves and polarization measurements, the data points in the time series are not independent. For blazar light curves, this is explained by the presence of flares, with gradual increases and decreases in brightness. The measured value at any given time depends on previous values. This behavior can be described by the DRW model, where the parameter $\tau_{DRW}$, also known as the damping time, reflects the "memory" of the process.

\subsubsection{Damped Random Walk}
\label{drw}
The DRW process, also known as the Ornstein-Uhlenbeck process, has gained popularity in the analysis of astronomical data over the last decade~\citep{ryan_2018, burke_2021, yang_2021, covino_2022, zhang_2022}. 

\cite{kelly_2009} have proposed to describe the optical variability of AGN as a random walk, which characterizes their light curves as a gaussian stochastic process with an exponential covariance function
\begin{equation}\label{eq_drw}
S(\Delta t) = \sigma_{DRW}^2 \exp\left(-\left|\frac{\Delta t}{\tau_{DRW}}\right|\right),
\end{equation}
which has two parameters: amplitude of variations $\sigma_{DRW}$ and the characteristic damping time $\tau_{DRW}$.

In some cases, deviations from the DRW model are observed for real object light curves. However, stochastic process models with a larger number of parameters, such as those assuming variation of an additional parameter to estimate the slope of the Structure Function (SF), as well as models that combine different stochastic processes, do not provide a better fit than a single DRW process~\citep{kelly_2011, zu_2013}.

The works of~\cite{kelly_2009, kozlowski_2010, macleod_2010} have shown that the DRW model provides a reasonable explanation for the variability of individual quasars in the Optical Gravitational Lensing Experiment (OGLE)~\citep{udalski_1997} and Sloan Digital Sky Survey (SDSS) Stripe 82 (S82)~\citep{sesar_2007} samples, and that the parameters $\tau_{DRW}$ and $\sigma_{DRW}$ correlate with such physical properties as luminosity and black hole mass. The study of~\cite{macleod_2011} demonstrated that the structure functions of quasars can be recovered by DRW modeling of light curves using estimates of the real physical parameters of the object. Furthermore, the DRW description is used for modeling AGN light curves in the task of reverberation mapping~\citep{zu_2011}. \cite{zhang_2022} and~\cite{zhang_2023} have also shown that this model is reliable for describing the variability of optical, X-ray, and gamma-ray emissions of blazars.

However, this model is not fully universal and has specific features that must be taken into account. The DRW model does not always provide a good fit between real and modeled data~\citep{kelly_2011}. The significance of deviations depends on data uncertainties. Deviations may be significant if the error estimation accuracy is less than $10\%$ and if the error distribution exhibits heavier tails than a normal distribution~\citep{zu_2013}. Moreover, studies~\cite{kozlowski_2010} and~\cite{burke_2021} report that the parameter $\tau_{DRW}$ may be incorrectly estimated for data sets that are too short or contain substantial gaps. Similar conclusions are drawn by~\cite{zu_2013}. \cite{burke_2021} argue that the estimate of $\tau_{DRW}$ can be considered reliable if it is greater than the mean cadence of the observations and less than 1/10 of the total duration of the observed light curve. \cite{kozlowski_2010} and~\cite{kozlowski_2016a} noted that errors in estimating $\tau_{DRW}$ are not critical if parameter tuning is not the main goal. This applies when the DRW model is used only to reconstruct or simulate data, not to estimate physical parameters.

In the current study we are modeling optical light curves of the blazars. For this purpose, we use the \textsc{javelin}\footnote{\url{https://github.com/nye17/javelin}} package. It was originally designed for reverbation mapping problems and utilizes the DRW process to model blazars light curves. The authors provide a detailed justification for the use of this approach in their works~\cite{zu_2011, zu_2013}.

It is important to note that stationarity is also required here, which may be violated in the case of long blazar's light curves. However, modeling non-stationary processes with uneven time sampling is considerably more complex, and therefore, we are compelled to make this assumption.

\subsubsection{Confidence Intervals Construction}
A comprehensive description of cross-correlation requires an assessment of its statistical significance. Internal correlations between adjacent points in the time series must be taken into account. Correlation measures the linear statistical relationship but does not imply causality between events. For example, when analyzing the correlation between two time series, each showing a flare, the cross-correlation function will exhibit a peak at some time lag, even if the signals are not related. 

\begin{figure}[ht!]
    \centering
    \includegraphics[width=\textwidth, angle=0]{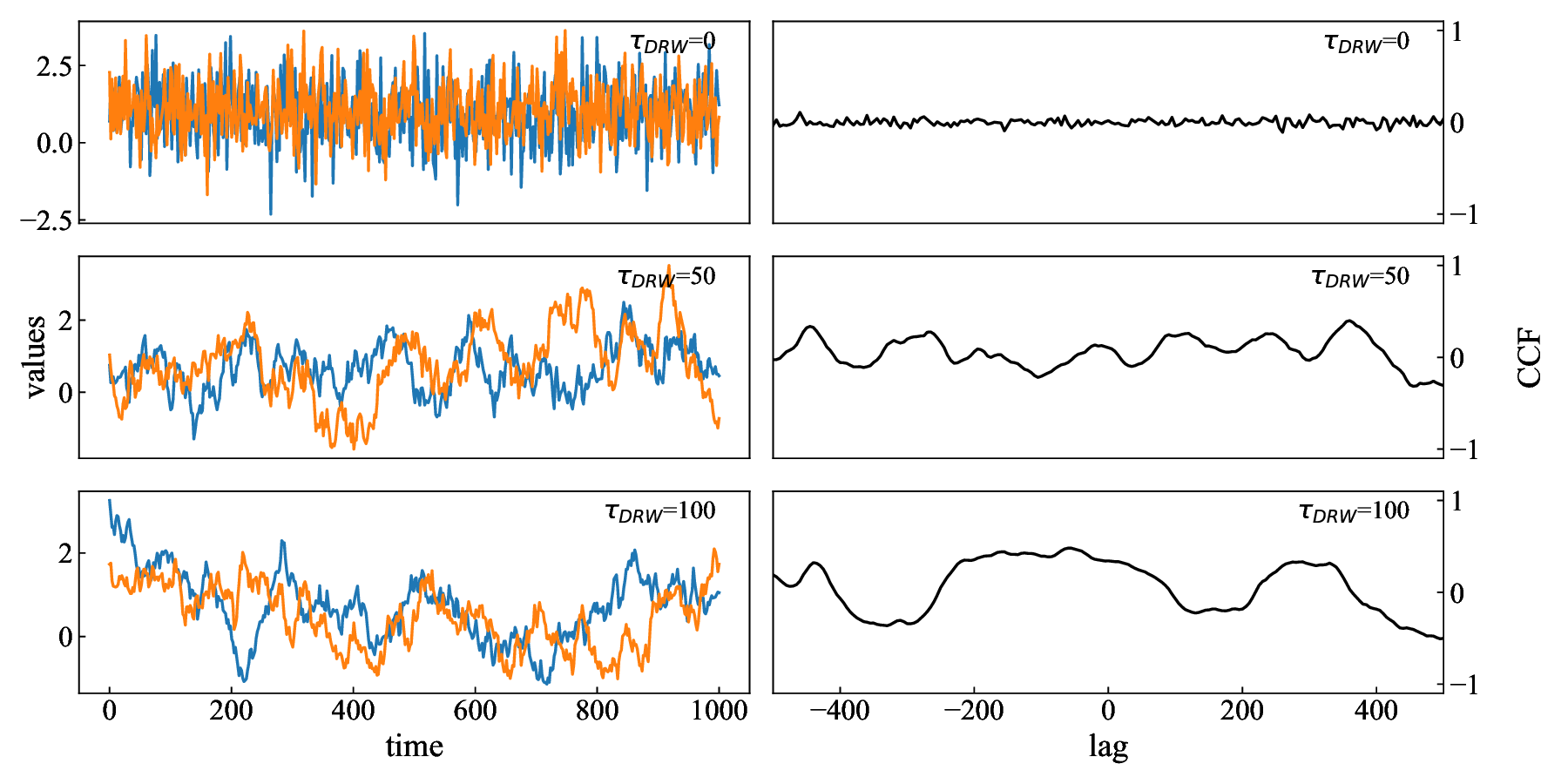}
    \caption{CCF (right) of two independent signals (left) corresponding to the DRW model with different $\tau_{DRW}$. The signals are defined on a uniform grid without the addition of random error.}
    \label{fig:drw_ccf}
\end{figure}

Standard methods commonly used to assess the significance of correlation assume that individual data points are uncorrelated. In the context of the DRW model, this behavior corresponds to the limiting case of $\tau_{DRW}=0$. However, for the vast majority of real curves, the $\tau_{DRW}$ parameter is significantly different from 0. Ignoring the internal correlation of the data leads to an overestimation of the significance of cross-correlations and erroneous physical interpretations. This effect is demonstrated in
Figure~\ref{fig:drw_ccf}, where for three simulated pairs of uncorrelated signals with different $\tau_{DRW}$ values the cross-correlation function is computed. It can be seen, that for signals with higher values of $\tau_{DRW}$ some spurious correlation peaks appear.

\begin{figure}[htbp]
    \centering
    \includegraphics[width=\textwidth]{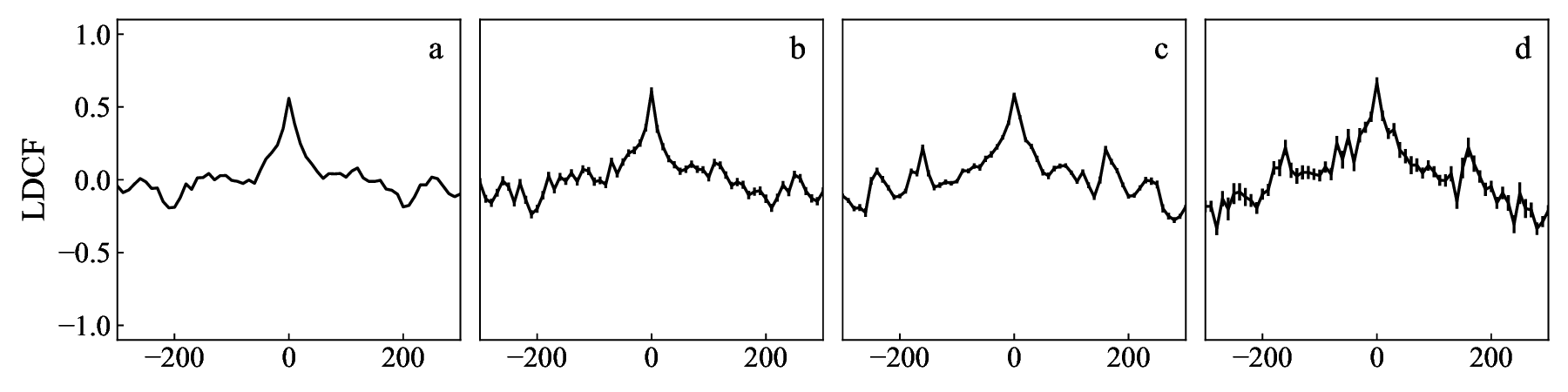}
    \caption{CCF estimation of two independent signals using LDCF method. (a) signals are generated without random errors, on a uniform grid; (b) signals are generated with random errors, on an irregular grid without significant gaps; (c) signals are generated with random errors, on a grid with large seasonal gaps; (d) signals are generated with random errors, on a grid with large seasonal gaps and additional irregular gaps.}
    \label{fig:grid}
\end{figure}

A reasonable approach is to use Monte Carlo simulations. Simulating artificial signals and calculating their cross-correlation on the actual time grid with real errors allows for both signal characteristics and observational imperfections to be accounted for. This is particularly important in the context of the LDCF method, as the choice of averaging interval width is critically important. If the interval contains too few points for proper averaging, it can lead to inflated/deflated false correlations. Below, Figure~\ref{fig:grid} illustrates the impact of grid irregularities and measurement errors on the cross-correlation estimation of two time series using the LDCF method. The signals are generated to exhibit identical behavior at $\text{lag} = 0$ and have a parameter $\tau_{DRW} = 25$. Gaps and errors were introduced randomly in this case solely to demonstrate the effect. In actual observations, such effects may become even more pronounced.

The idea of using Monte Carlo simulations in the context of astronomical data analysis has been discussed and applied by other authors~\citep{edelson_1995, uttley_2003,  moerbeck_2014}. The details of the procedure may vary from study to study. Typically, methods based on power spectrum density or structure function are used to model AGN light curves. However, as noted in~\cite{emmanoulopoulos_2010}, using structure functions to assess blazar variability comes with several challenges and limitations. The method based on PSD, first proposed in~\cite{uttley_2003} and improved in~\cite{moerbeck_2014}, is more justified but has its downsides. The improved method involves transforming into the frequency domain, which requires complex mathematical transformations and the use of window functions. Moreover, to obtain a smooth periodogram, additional filtering and interpolation in the time domain are necessary. The final technique becomes complex to understand and reproduce. When simulating synthetic curves to construct confidence intervals using such a model, the result depends not only on grid discretization, curve length, and measurement errors but also on the choice of window function.

All these complexities are justified in the context of the challenging task of correlational analysis of irregular time series. However, we aim to find a simpler, more intuitive method that requires fewer manipulations and transformations. Therefore, we explore the possibility of using the DRW model for simulating artificial curves. This approach does not require prior interpolation or data smoothing, nor does it involve transitioning from the time domain to the frequency domain.

The overall scheme of our procedure for estimating cross-correlations and constructing confidence intervals (CIs) is shown below. It can be applied to any curves with an appropriate distribution of data points, e.g., for two light curves. In our case, these are the light (L) and the polarization degree curves (PD).
\begin{enumerate}
    \item Calculate the cross-correlations of the uneven time series (L and PD) using the selected method.
    \item Estimate the parameters $\tau_{DRW}$ and $\sigma_{DRW}$ of the model for one of the analyzed curves (L). 
    \item Simulate $N$ artificial curves based on the DRW model with the parameters estimated in the previous step and add random errors corresponding to the measurement errors. The artificial (L) curves are defined on the same time grid as the original data. 
    \item Calculate the cross-correlations of the (PD) curve, for which steps 2-3 were not performed, with each of the $N$ artificial (L) curves. For each pair, the values of the cross-correlation at the corresponding time delay $\tau$ are recorded.
    \item Assess the significance levels of the cross-correlation coefficients of the original data based on the empirical distribution of the cross-correlation values at each time delay $\tau$. The $2\sigma$ and $3\sigma$ significance levels correspond to the $95^{th}$ and $99^{th}$ percentiles of the obtained distributions.
\end{enumerate}
To estimate the confidence intervals (CIs), we propose simulating only the optical light curves. Since the \textsc{javelin} library we are using was originally developed for optical reverberation mapping, we lack an appropriate tool for modeling polarization degree curves. For these reasons, we focus exclusively on modeling artificial optical light curves. This choice also significantly reduces computation time without violating the conditions of the task. Our approach evaluates the probability that the observed cross-correlation pattern arises by chance, given the current characteristics of the data. The artificial series, modeled using the parameters of one curve, preserve its statistical properties, while the second curve retains its original characteristics.

\section{Data analysis}
\label{sect:analysis}
The main goal of our research is to analyze the correlations between long-term observational series of optical brightness and polarization degree, corresponding to $lag\approx0$. In other words, we are interested in identifying objects for which the behaviors of these characteristics are correlated and occur simultaneously or quasi-simultaneously, and whether this correlated behavior is associated with the types of the studied objects. 

The cross-correlation was evaluated using the LDCF algorithm. Our choice to conduct cross-correlation analysis specifically on long-term light curves is deliberate. Although some authors report correlations between light curve behavior and optical polarization only during periods of high activity~\citep{pandey_2021, bachev_2023}, this study focuses on the overall analysis of the curves. Such an approach allows for a more reliable reconstruction of the DRW model parameters (Sec.~\ref{drw}) and the use of DRW curve simulations to estimate CIs.

As previously mentioned, we employed the \textsc{javelin} package for modeling the optical light curves of blazars. Parameter fitting employs a Markov Chain Monte Carlo (MCMC) sampler via the \textsc{emcee}\footnote{\url{https://emcee.readthedocs.io/en/stable/}} package. Further details can be found in~\cite{zu_2011, zu_2013} and the library documentation.

When estimating the parameters of the optical light curves, a uniform distribution is used by default to estimate $\tau_{DRW}$  and $\sigma_{DRW}$ in \textsc{javelin}, but with additional constraints. For both parameters, a uniform distribution with a logarithmic scale is assumed. For $\sigma_{DRW}$, a penalty is applied by calculating $-log(\sigma)$. For $\tau_{DRW}$, the penalty is adjusted as follows: if $\tau_{DRW}$ exceeds the duration $D$ of the data, the penalty is calculated as $-log(\frac{\tau}{D})$. If $\tau_{DRW}$ is less than $0.001$, a penalty of $-\infty$ is applied (i.e., this value is considered completely unacceptable). The MCMC sampler was run for $10^{5}$ iterations with $16$ parallel walkers exploring the parameter space. The first $3 \cdot 10^{3}$ steps were used as burn-in. The result of this fitting process is an estimate of the posterior distribution of the parameters along with corresponding highest posterior density intervals. The parameter $\tau_{DRW}$ was compared with the limits proposed in~\cite{burke_2021}: it should be no smaller than the mean cadence and no larger than $1/10$ of the total length of the studied curve.

After parameter estimation, around $10^{5}$ artificial light curves were generated to evaluate the significance of the cross-correlation of the original signals. The number of artificial light curves was set depending on the convergence and computational cost of the generation and confidence interval calculation algorithms. Convergence was assessed by the appearance of the empirical distribution of cross-correlation coefficients, only for the zero time lag.

\section{Results}
\label{sect:results}
Below are the results of the cross-correlation values computed using the LDCF method and the
estimation of CI for several illustrative cases. The results of cross-correlation evaluations and the parameters of the optical brightness curves $\sigma_{DRW}$ and $\tau_{DRW}$ for all objects are provided in Table~\ref{tab:results}.

\subsection{OJ 248}
\begin{figure}[ht!]
    \begin{minipage}{0.65\textwidth}
        \centering
        \includegraphics[width=\textwidth]{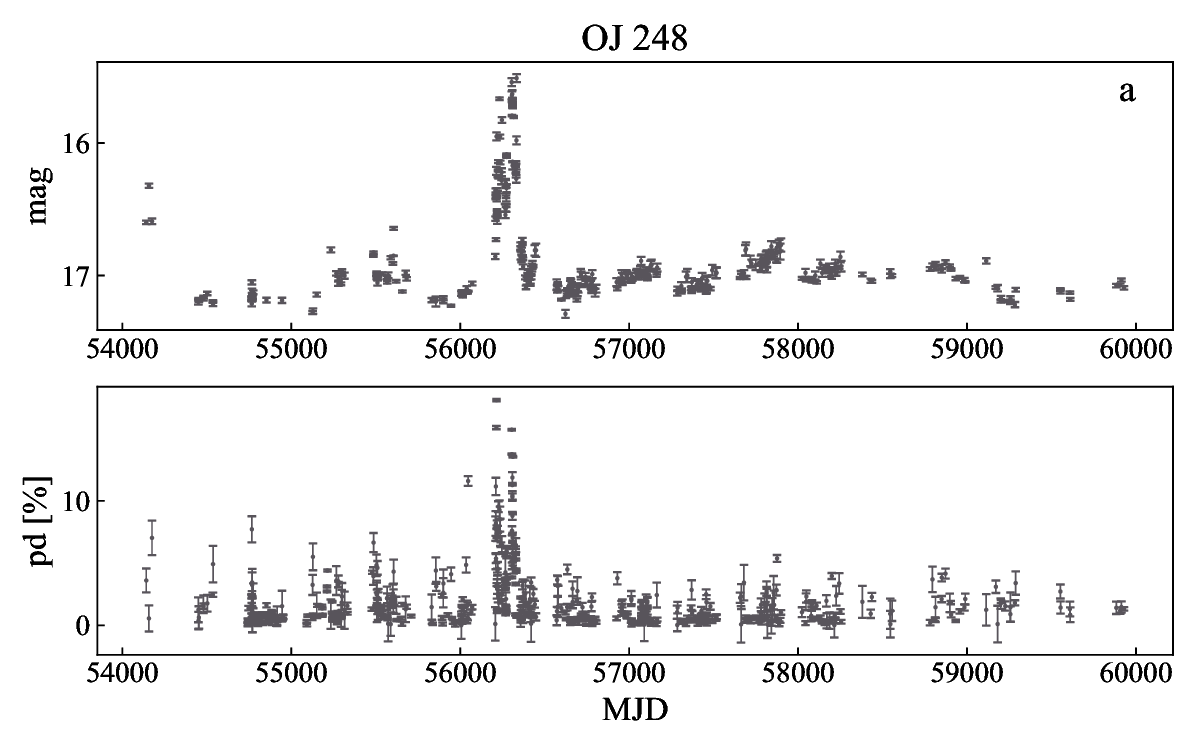}
    \end{minipage}
    \hspace{0mm}    
    \begin{minipage}{0.4\textwidth}
        \centering
        \includegraphics[width=\textwidth]{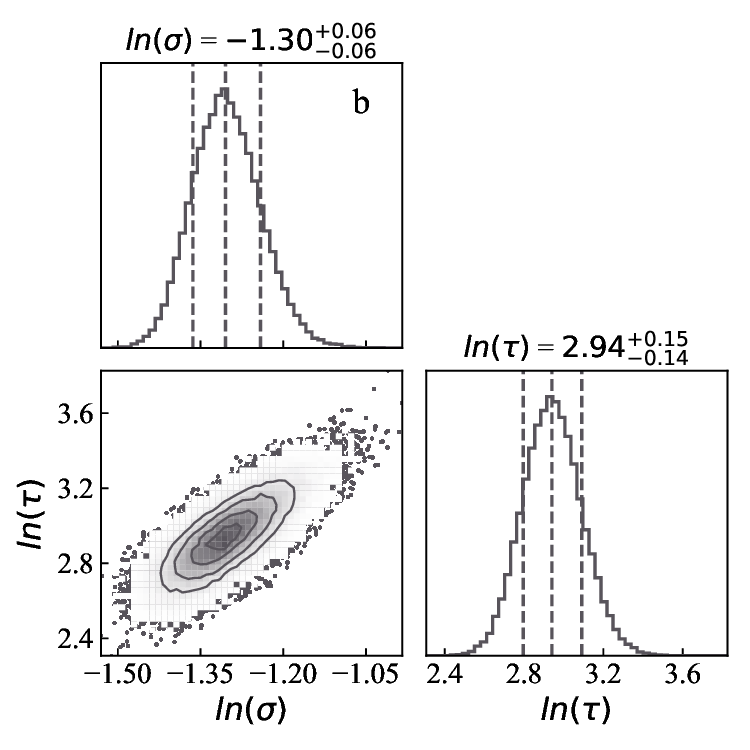}
    \end{minipage}
    
    \begin{minipage}{\textwidth}
        \centering
        \includegraphics[width=\textwidth]{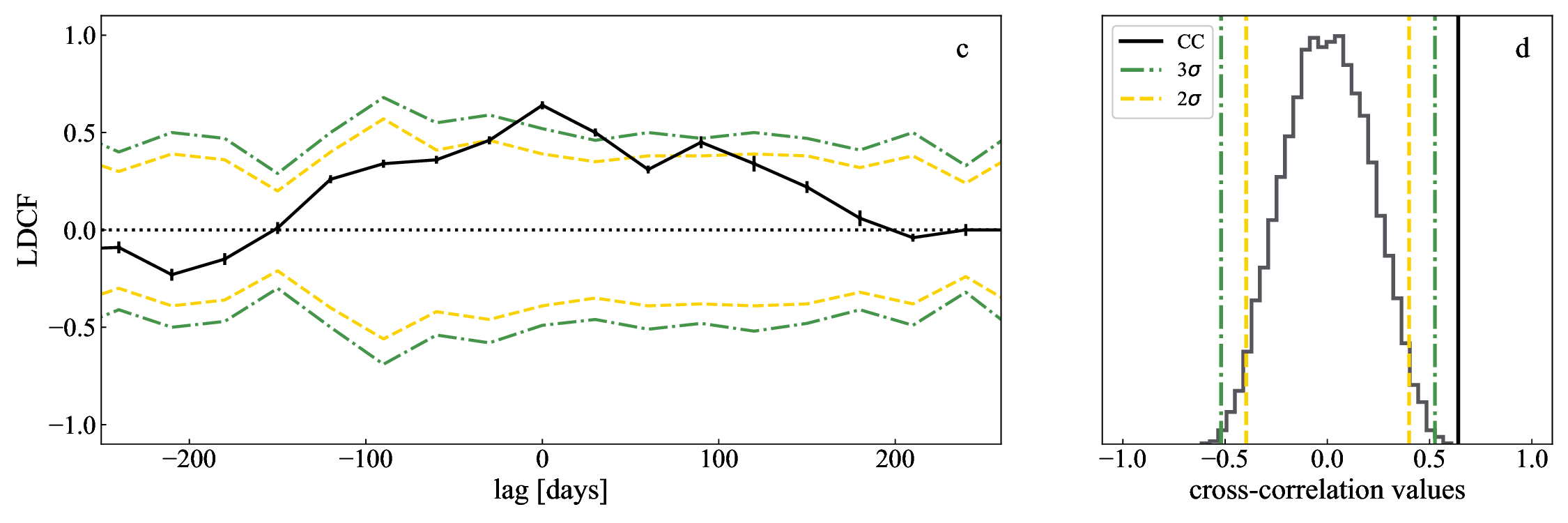}
    \end{minipage}
    
    \caption{OJ 248. (a) optical \textit{R} and PD curves; (b) posterior distribution estimates for $\sigma_{DRW}$ and $\tau_{DRW}$ for \textit{R} curve; (c) CCF estimation results with CI evaluation; (d) empirical distribution of cross-correlation values obtained from MC simulations for $lag=0$.}
    \label{fig:oj248}
\end{figure}

The light curves of the FSRQ object OJ 248 (Fig.~\ref{fig:oj248}) provide a good example of the correlated behavior of the polarization degree and optical brightness. The characteristics exhibit similar behavior, both showing a flare localized in 56000-56500 MJD. The correlation is confirmed by our evaluation --- a significant CCF peak is observed at $lag=0$. The asymmetries of the CIs for this and other cases are due to overall irregularities, seasonal gaps, and the length of the observed curves. This leads to different numbers of point pairs in each averaging interval, which is reflected in the irregular shape of the CI. The estimates of the $\tau_{DRW}$ parameter for \textit{R} curve are consistent with the limits proposed in~\cite{burke_2021}. A correlation between the optical light curve and degree of polarization curve for this source was also reported by~\cite{carnerero_2015},~\cite{raiteri_2021} and~\cite{otero_2023}.

\subsection{W Com}
\begin{figure}[ht!]
    \begin{minipage}{0.65\textwidth}
        \centering
        \includegraphics[width=\textwidth]{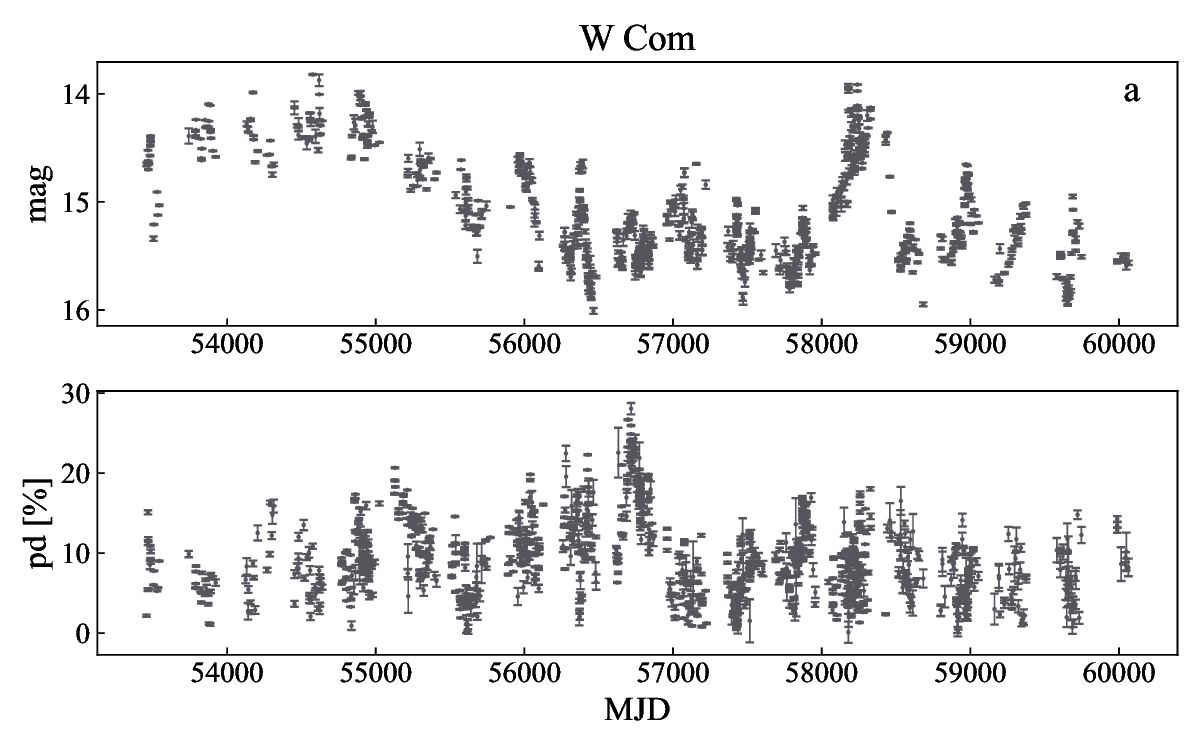}
    \end{minipage}
    \hspace{0mm}    
    \begin{minipage}{0.4\textwidth}
        \centering
        \includegraphics[width=\textwidth]{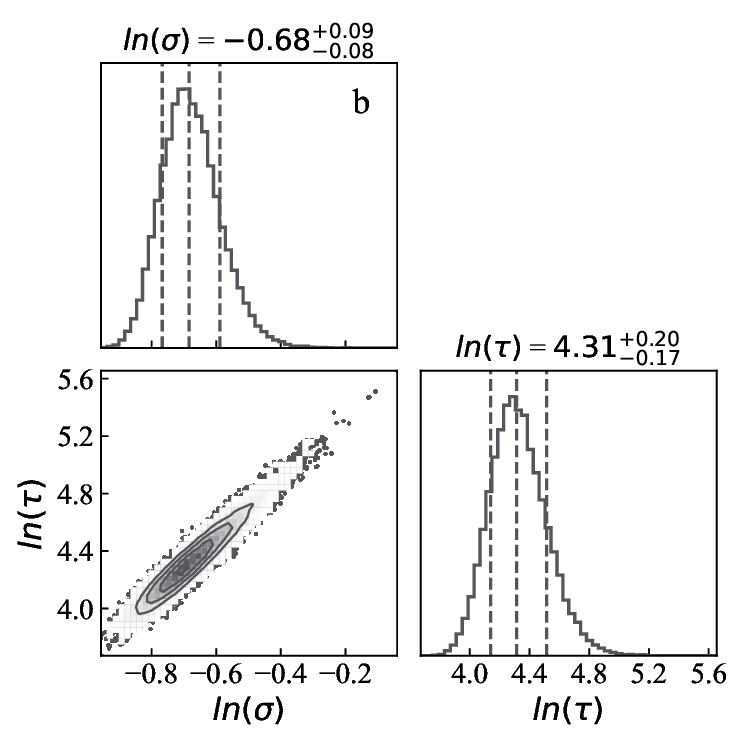}
    \end{minipage}
    
    \begin{minipage}{\textwidth}
        \centering
        \includegraphics[width=\textwidth]{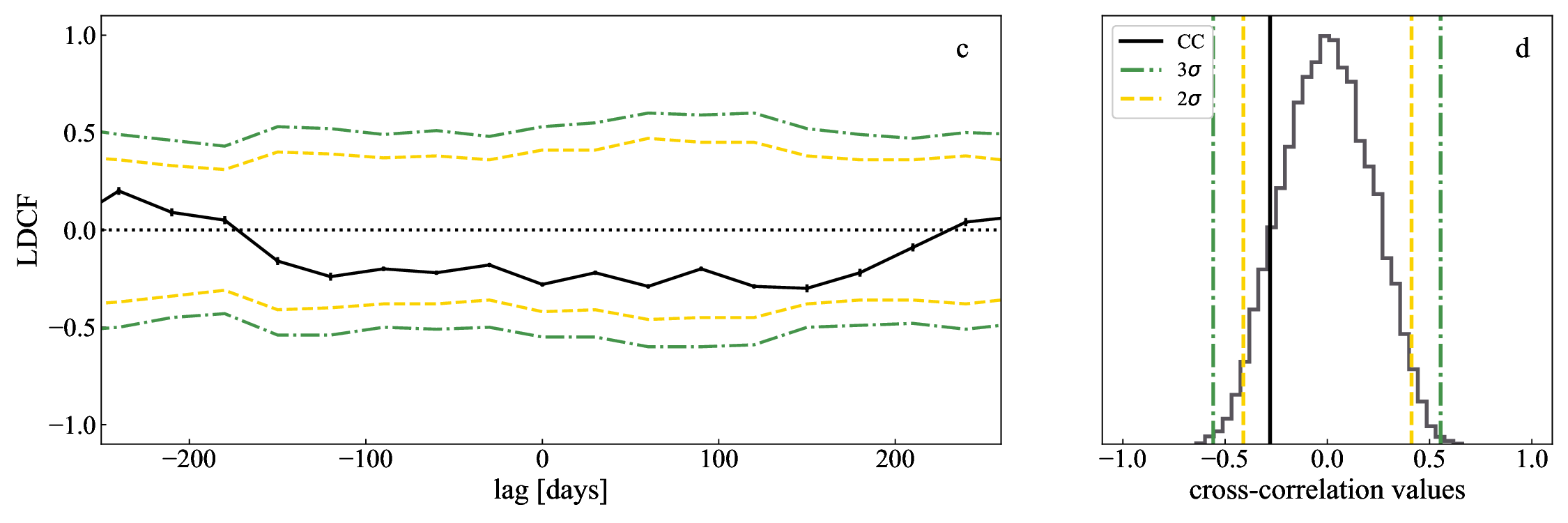}
    \end{minipage}
    
    \caption{W Com. (a) optical \textit{R} and PD curves; (b) posterior distribution estimates for $\sigma_{DRW}$ and $\tau_{DRW}$ for \textit{R} curve; (c) CCF estimation results with CI evaluation; (d) empirical distribution of cross-correlation values obtained from MC simulations for $lag=0$.}
    \label{fig:wcom}
\end{figure}

The curves of this BL~Lac object (see Fig.~\ref{fig:wcom}) exhibit uncorrelated behavior between optical brightness and the polarization degree, which is confirmed by our CCF estimation. The estimates of the $\tau_{DRW}$ parameter are consistent with the bounds proposed in the study~\cite{burke_2021}.

% W Com

\subsection{3C 84}
\begin{figure}[ht!]
    \begin{minipage}{0.65\textwidth}
        \centering
        \includegraphics[width=\textwidth]{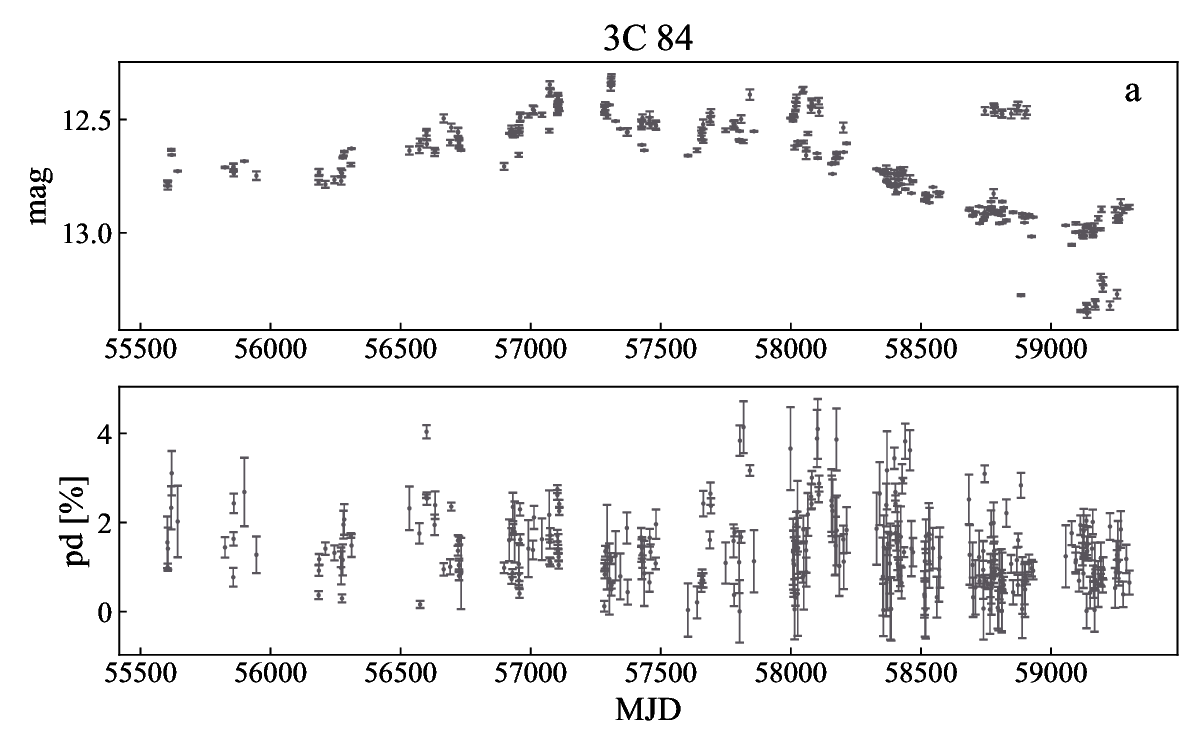}
    \end{minipage}
    \hspace{0mm}    
    \begin{minipage}{0.4\textwidth}
        \centering
        \includegraphics[width=\textwidth]{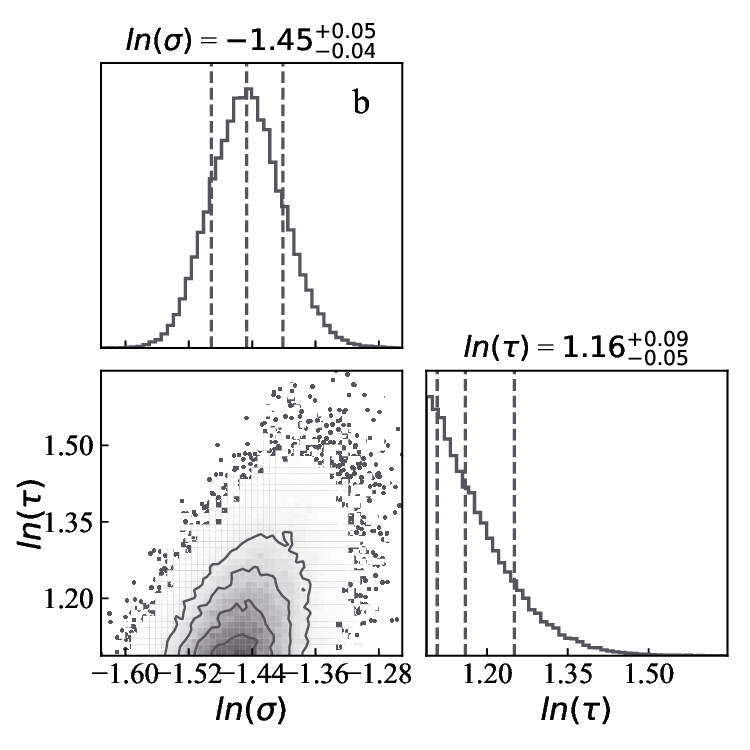}
    \end{minipage}
    
    \begin{minipage}{\textwidth}
        \centering
        \includegraphics[width=\textwidth]{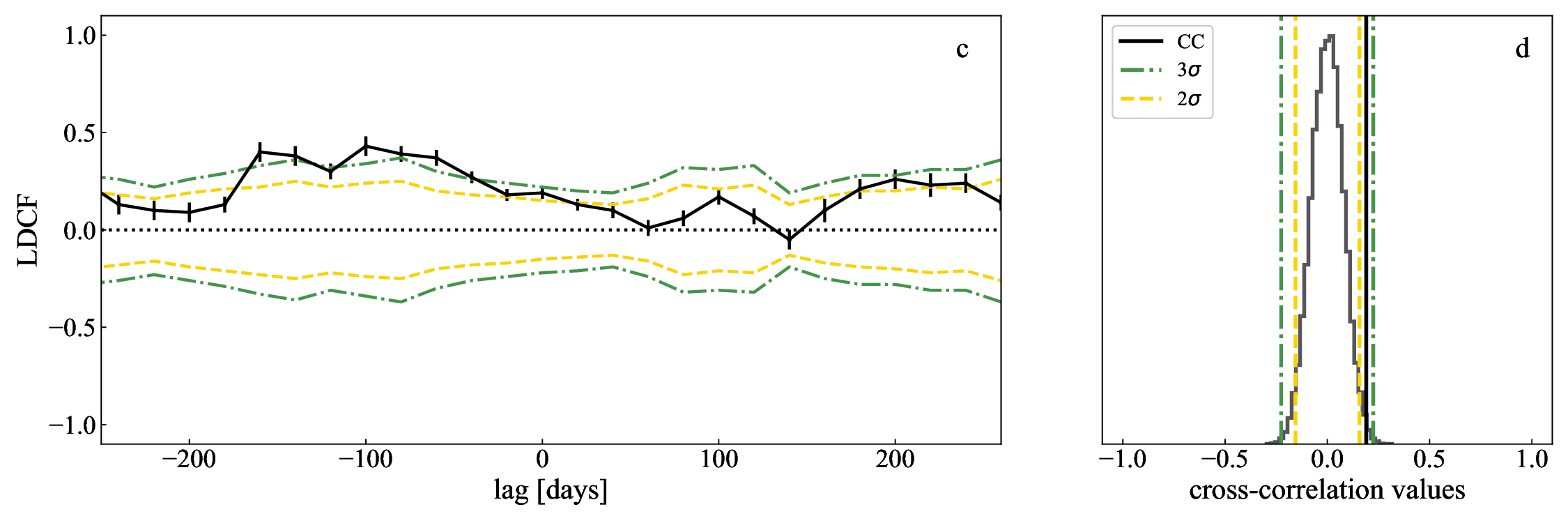}
    \end{minipage}
    
    \caption{3C 84. (a) optical \textit{R} and PD curves; (b) posterior distribution estimates for $\sigma_{DRW}$ and $\tau_{DRW}$ for \textit{R} curve; (c) CCF estimation results with CI evaluation; (d) empirical distribution of cross-correlation values obtained from MC simulations for $lag=0$.}
    \label{fig:3c84}
\end{figure}

For this radio galaxy, we were unable to effectively estimate the $\tau_\mathrm{DRW}$ parameter for the light curve or construct a CI estimate (Fig.~\ref{fig:3c84}). The resulting estimate of $\tau_\mathrm{DRW}$ is outside the bounds proposed in~\cite{burke_2021}. The CI limits are too narrow, as if the process describing this object's light curve were almost memory-less. We attribute this to the small number of data points in the light curve and the large measurement errors in the polarization degree. We were unable to improve the parameter estimate by adjusting the settings of the \textsc{javelin} package we used. Unfortunately, the functionality of this library is limited, as it was primarily developed for the task of reverberation mapping. We hope to use a more flexible tool in future studies.

% % 3C 84

\subsection{AO 0235+16}
\begin{figure}[ht!]
    \begin{minipage}{0.65\textwidth}
        \centering
        \includegraphics[width=\textwidth]{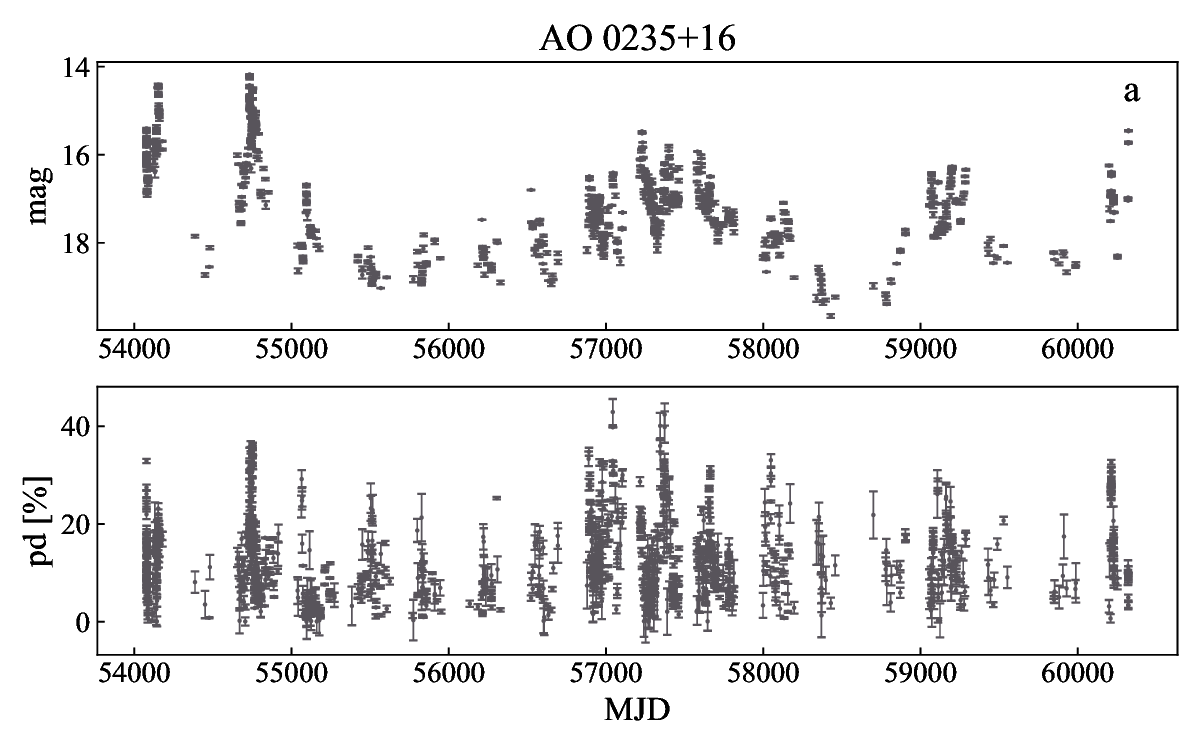}
    \end{minipage}
    \hspace{0mm}    
    \begin{minipage}{0.4\textwidth}
        \centering
        \includegraphics[width=\textwidth]{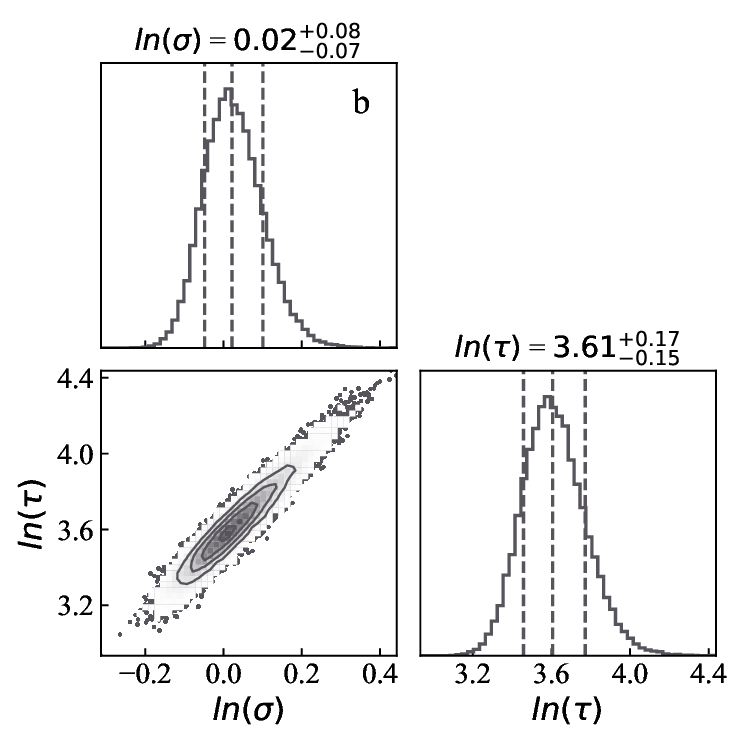}
    \end{minipage}
    
    \begin{minipage}{\textwidth}
        \centering
        \includegraphics[width=\textwidth]{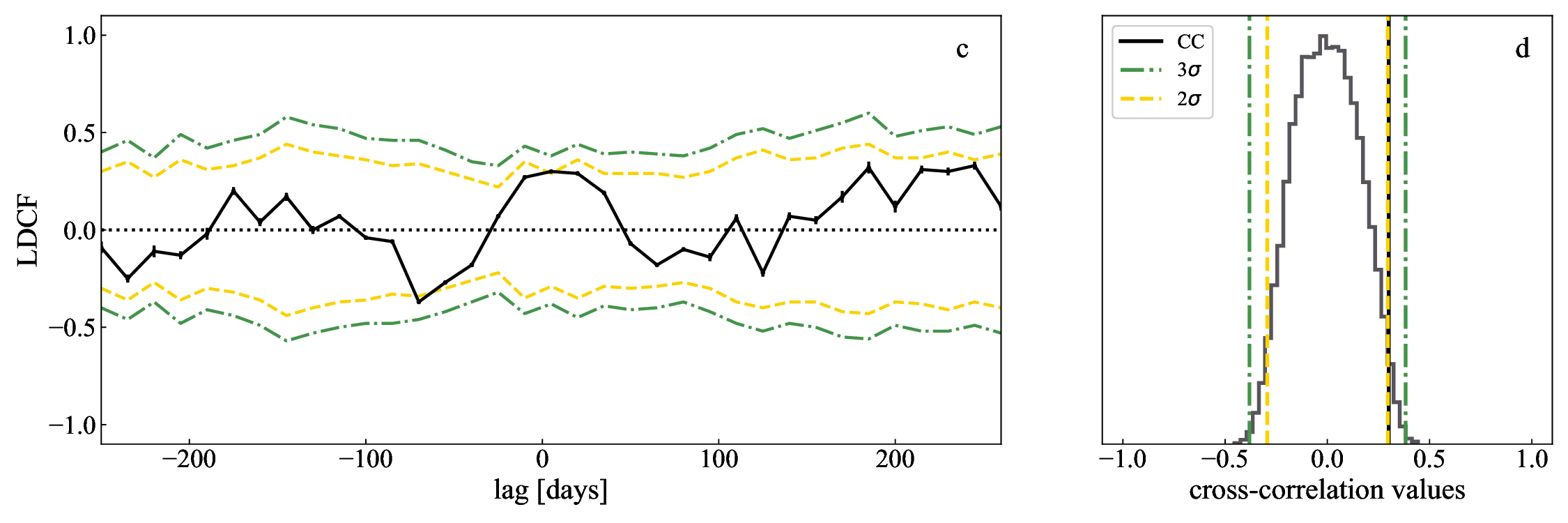}
    \end{minipage}
    
    \caption{AO 0235+16. (a) optical \textit{R} and PD curves; (b) posterior distribution estimates for $\sigma_{DRW}$ and $\tau_{DRW}$ for \textit{R} curve; (c) CCF estimation results with CI evaluation; (d) empirical distribution of cross-correlation values obtained from MC simulations for $lag=0$.}
    \label{fig:ao0235}
\end{figure}

This case illustrates the conservativeness of the cross-correlation estimates we obtained (Fig.~\ref{fig:ao0235}). For this FSRQ object the cross-correlation estimate at zero lag barely exceed the computed confidence intervals, despite the visually similar behavior of the curves. It is possible that with denser observation series with smaller errors, the correlation could have been confirmed more clearly. The estimates of the $\tau_{DRW}$ parameter for \textit{R} curve are in agreement with the limits proposed in \cite{burke_2021}.

% AO 0235+16

\subsection{Other Results}
All results are presented in Table~\ref{tab:results}. For 6 objects, we were able to confirm the correlation in the changes of optical brightness and polarization degree. Among them, 5 are FSRQ objects and 1 is a BL Lac object. For 16 objects, we conclude the absence of correlation in the studied curves at zero lag: 10 BL~Lac and 6 FSRQ. For the radio galaxy 3C 84, we were unable to obtain a reliable estimate. For PKS~0420-01 and for AO~0235+16, which show visually similar behavior in their curves, the cross-correlation estimates barely exceed $2\sigma$ threshold. The result depends on the quantity and density of available observations -- for some sources, we may not have sufficient data to reliably characterize the behavior of the studied light curves. Additionally, for the Mkn~421, a noticeable CCF peak exceeding $3\sigma$ was observed at $lag \approx 50\,$days. However, we cannot be confident in its significance, as the empirical distribution of cross-correlation coefficients at this lag was not studied. No significant anticorrelation was observed in any of the investigated cases. 

\begin{table}[ht!]
\begin{center}
\caption[]{Results.}\label{tab:results}
\begin{tabular}{lccc}
\hline\noalign{\smallskip}
Name & Correlation & $\tau_{DRW}$  & $\sigma_{DRW}$ \\
& (Significance) & (days) &   \\
\hline\noalign{\smallskip}
3C 66A & N & $102.68^{+126.88}_{-86.03}$ & $0.44^{+0.48}_{-0.4}$ \\ 
OJ 049 & N & $16.52^{+19.53}_{-14.21}$ & $0.46^{+0.49}_{-0.43}$ \\ 
Mkn 421 & N & $14.65^{+16.34}_{-13.21}$ & $0.4^{+0.42}_{-0.38}$ \\ 
S4 0954+65 & Y $(2\sigma)$ & $14.8^{+16.19}_{-13.61}$ & $0.88^{+0.92}_{-0.85}$ \\ 
BL Lacertae & N & $30.56^{+34.16}_{-27.83}$ & $0.65^{+0.68}_{-0.62}$ \\ 
OJ 287 & N & $38.14^{+43.81}_{-33.78}$ & $0.53^{+0.57}_{-0.5}$ \\ 
PG 1553 & N & $118.21^{+167.77}_{-89.42}$ & $0.28^{+0.33}_{-0.25}$ \\ 
PKS 0735+17 & N & $72.83^{+89.1}_{-60.55}$ & $0.59^{+0.65}_{-0.54}$ \\ 
Q 1959 & N & $79.02^{+98.91}_{-65.24}$ & $0.32^{+0.35}_{-0.29}$ \\ 
S5 0716+71 & N & $15.53^{+16.83}_{-14.42}$ & $0.55^{+0.58}_{-0.53}$ \\ 
W Com & N & $74.69^{+91.33}_{-62.78}$ & $0.5^{+0.55}_{-0.46}$ \\ 
AO 0235+16 & Y $(2\sigma)$ & $36.82^{+43.46}_{-31.76}$ & $1.02^{+1.11}_{-0.95}$ \\ 
3C 454.3 & Y $(2\sigma)$ & $39.46^{+45.68}_{-34.57}$ & $0.73^{+0.79}_{-0.69}$ \\ 
OJ 248 & Y $(3\sigma)$ & $18.93^{+22.0}_{-16.38}$ & $0.27^{+0.29}_{-0.26}$ \\ 
CTA 102 & Y $(3\sigma)$ & $33.98^{+38.63}_{-30.13}$ & $0.86^{+0.92}_{-0.81}$ \\ 
3C 273 & N & $240.16^{+350.6}_{-180.15}$ & $0.16^{+0.19}_{-0.14}$ \\ 
3C 279 & N & $47.02^{+55.89}_{-40.4}$ & $0.82^{+0.89}_{-0.76}$ \\ 
CTA 26 & N & $33.49^{+40.64}_{-28.01}$ & $0.66^{+0.72}_{-0.61}$ \\ 
PKS 0420-01 & Y $(2\sigma)$ & $88.7^{+112.49}_{-72.33}$ & $0.96^{+1.07}_{-0.87}$ \\ 
Q 0836 & N & $13.47^{+16.16}_{-11.28}$ & $0.09^{+0.1}_{-0.09}$ \\ 
Q 1156 & N & $54.21^{+64.04}_{-46.71}$ & $1.14^{+1.23}_{-1.06}$ \\ 
PKS 1222+21 & N & $145.93^{+196.65}_{-114.93}$ & $0.43^{+0.49}_{-0.38}$ \\ 
3C 84 & ? & $3.19^{+3.49}_{-3.03}$ & $0.24^{+0.25}_{-0.23}$ \\
\noalign{\smallskip}\hline
\end{tabular}
\end{center}
\end{table}

\section{Discussion and Conclusions}
\label{sect:discussion}
We have compared our results with the finding by~\cite{otero_2023}, which also focuses on the analysis of long-term blazar observations but uses different methods. The authors report, before correcting the polarization degree for the host galaxy and BLR, a correlation between optical flux and polarization degree observed for 5 of 11 FSRQ objects in their sample, while no correlation is found for BL~Lac objects. The sources which reveal a correlated behavior include 4 FSRQs from our study: PKS~0420-01, OJ~248, 3C~454.3, and CTA~102. 

The results of our study can be called in agreement with the findings of~\cite{otero_2023}. In our sample, 5 out of 6 objects showing significant correlation at zero lag are FSRQ objects. Although, this is not enough to draw definitive conclusions. To test whether the difference in the correlated behavior of the curves between the two types of blazars is significant, we performed a two-sided Fisher's exact test for small binomial samples. The obtained $p-value\approx0.15$ exceeds the significance level $\alpha=0.05$, indicating insufficient evidence to reject the null hypothesis (i.e., the samples are equal). Therefore, based on the current data, there is no statistically significant evidence of differences in the emission characteristics between the two blazar classes.

It is possible that the observed correlation between optical flux and polarization degree does not necessarily reflect a stable relationship between these characteristics over the entire observation period, but is more likely related to the high correlation during individual outbursts. FSRQ objects generally exhibit brighter and more frequent flares compared to BL~Lac objects.

The contribution of flares to the observed correlation is especially evident in the case of OJ 248, which demonstrates a single bright flare. For this object, a re-evaluation of the correlation was performed outside the flare state. It was found that the high correlation between examined characteristics disappears during quiescent periods. Similar high correlations for other objects may also be explained by the contribution of simultaneous flare events to the changing flux and polarization degree curves.

The correlation between flare appearances on the flux and polarization degree curves can be explained by shock-in-jet models~\citep{hagen_2008, marscher_1996}. In this model, shocks in the jet cause the injection of relativistic electrons into the emission region, leading to a stable spectral shape and short-term variability in the optical range. Furthermore, shocks contribute to the alignment of the magnetic field, resulting in a positive correlation between flux and polarization degree. A similar model is also discussed by~\cite{angelakis_2016}, describing the differences in the observed optical emission for low-synchrotron peaked (LSP) and high-synchrotron peaked (HSP) objects. Since the optical emission for FSRQ/LSP and ISP/HSP objects is located in different regions of the synchrotron emission curve in the spectral energy distribution, this model explains the higher polarization and variability for the former, compared to the latter. ~\cite{otero_2023} also proposed the shock-in-jet model as an explanation for the observed patterns. 

Finding more cases of a significant positive correlation between the optical light curve and degree of polarization curve in FSRQs can be a signature of more powerful shocks in quasars than in BL Lac objects. This assumption agrees also with the presumption that the magnetic field energy in BL Lac objects dominates the particle energy even on parsec scales, e.g. for BL Lac itself~\citep{cohen_2014}. For a better understanding of the mechanisms leading to the observed features of the emission and the selection of the most appropriate model, further investigation of other characteristics of polarized emission is required. 

We also discussed a method for assessing the significance of cross-correlation in time series, focusing on model selection and problem formulation. Our goal was to find a simple, efficient, and universal approach for evaluating the cross-correlation of uneven time series with various characteristics. While the combination of cross-correlation estimation and confidence interval methods we used performs well, it is computationally demanding and may not be applicable in all cases due to the complexity of the problem. Nevertheless, these methods provide accurate estimates, and we hope to refine the approach and compare it with other correlation estimation and CI calculation methods in the future.

\begin{acknowledgements}
This study was based in part on observations conducted using the $1.8\,$m Perkins Telescope Observatory in Arizona, which is owned and operated by Boston University. 
The research at Boston University was supported in part by the National Science Foundation grant AST-2108622, and several NASA Fermi Guest Investigator grants, the latest is 80NSSC23K1507.

Data from the Steward Observatory spectropolarimetric monitoring project were also used. This program is supported by Fermi Guest Investigator grants NNX08AW56G, NNX09AU10G, NNX12AO93G, and NNX15AU81G. 

The authors express their gratitude to the anonymous reviewer for comments that helped improve this work.
\end{acknowledgements}

\bibliographystyle{raa}
\bibliography{ms2025-0051main}

\begin{thebibliography}{54}
\providecommand\natexlab[1]{#1}
\providecommand\JournalTitle[1]{#1}

\bibitem[Angelakis {et~al.}(2016)]{angelakis_2016}
Angelakis, E., Hovatta, T., Blinov, D., {et~al.} 2016, Monthly Notices of the Royal Astronomical Society, 463, 3365

\bibitem[Bachev {et~al.}(2023)]{bachev_2023}
Bachev, R., Tripathi, T., Gupta, A.~C., {et~al.} 2023, Monthly Notices of the Royal Astronomical Society, 522, 3018

\bibitem[Blandford {et~al.}(1978)]{blandford_1978}
Blandford, R.~D., Rees, M.~J., \& M., W.~A. 1978, Pittsburgh Conference on BL Lac Objects, University of Pittsburgh, Pittsburgh, Pa., April 24-26, 1978, Proceedings (Pittsburgh: University of Pittsburgh)

\bibitem[{Burke} {et~al.}(2021)]{burke_2021}
{Burke}, C.~J., {Shen}, Y., {Blaes}, O., {et~al.} 2021, Science, 373, 789

\bibitem[Carnerero {et~al.}(2015)]{carnerero_2015}
Carnerero, M.~I., Raiteri, C.~M., Villata, M., {et~al.} 2015, \mnras, 450, 2677

\bibitem[{Chen} {et~al.}(2024)]{chen_2024}
{Chen}, Y., {Gu}, Q., {Yang}, J., {et~al.} 2024, Research in Astronomy and Astrophysics, 24, 115011

\bibitem[Cohen {et~al.}(2014)]{cohen_2014}
Cohen, M.~H., Meier, D.~L., Arshakian, T.~G., {et~al.} 2014, The Astrophysical Journal, 787, 151

\bibitem[Covino {et~al.}(2022)]{covino_2022}
Covino, S., Tobar, F., \& Treves, A. 2022, \mnras, 513, 2841

\bibitem[{Covino} {et~al.}(2015)]{covino_2015}
{Covino}, S., {Baglio}, M.~C., {Foschini}, L., {et~al.} 2015, \aap, 578, A68

\bibitem[{Edelson} \& {Krolik}(1988)]{edelson_1988}
{Edelson}, R.~A., \& {Krolik}, J.~H. 1988, \apj, 333, 646

\bibitem[{Edelson} {et~al.}(1995)]{edelson_1995}
{Edelson}, R., {Krolik}, J., {Madejski}, G., {et~al.} 1995, \apj, 438, 120

\bibitem[{Emmanoulopoulos} {et~al.}(2010)]{emmanoulopoulos_2010}
{Emmanoulopoulos}, D., {McHardy}, I.~M., \& {Uttley}, P. 2010, \mnras, 404, 931

\bibitem[{Fraija} {et~al.}(2017)]{fraija_2017}
{Fraija}, N., {Ben{\'\i}tez}, E., {Hiriart}, D., {et~al.} 2017, \apjs, 232, 7

\bibitem[{Hagen-Thorn} {et~al.}(2008)]{hagen_2008}
{Hagen-Thorn}, V.~A., {Larionov}, V.~M., {Jorstad}, S.~G., {et~al.} 2008, \apj, 672, 40

\bibitem[{Hufnagel} \& {Bregman}(1992)]{hufnagel_1992}
{Hufnagel}, B.~R., \& {Bregman}, J.~N. 1992, \apj, 386, 473

\bibitem[{Itoh} {et~al.}(2018)]{itoh_2018}
{Itoh}, R., {Uemura}, M., {Fukazawa}, Y., \& {Kawabata}, K. 2018, Galaxies, 6, 16

\bibitem[{Jorstad} {et~al.}(2010)]{jorstad_2010}
{Jorstad}, S.~G., {Marscher}, A.~P., {Larionov}, V.~M., {et~al.} 2010, \apj, 715, 362

\bibitem[{Kelly} {et~al.}(2009)]{kelly_2009}
{Kelly}, B.~C., {Bechtold}, J., \& {Siemiginowska}, A. 2009, \apj, 698, 895

\bibitem[{Kelly} {et~al.}(2011)]{kelly_2011}
{Kelly}, B.~C., {Sobolewska}, M., \& {Siemiginowska}, A. 2011, \apj, 730, 52

\bibitem[Koz{\l}owski(2016)]{kozlowski_2016a}
Koz{\l}owski, S. 2016, The Astrophysical Journal, 826, 118

\bibitem[{Koz{\l}owski} {et~al.}(2010)]{kozlowski_2010}
{Koz{\l}owski}, S., {Kochanek}, C.~S., {Udalski}, A., {et~al.} 2010, \apj, 708, 927

\bibitem[Laing(1980)]{laing_1980}
Laing, R.~A. 1980, Monthly Notices of the Royal Astronomical Society, 193, 439

\bibitem[{Larionov} {et~al.}(2008)]{larionov_2008}
{Larionov}, V.~M., {Jorstad}, S.~G., {Marscher}, A.~P., {et~al.} 2008, \aap, 492, 389

\bibitem[MacLeod {et~al.}(2011)]{macleod_2011}
MacLeod, C.~L., Brooks, K., Ivezi{\'c}, {\v{Z}}., {et~al.} 2011, The Astrophysical Journal, 728, 26

\bibitem[{MacLeod} {et~al.}(2010)]{macleod_2010}
{MacLeod}, C., {Ivezic}, Z., {Kozlowski}, S., {et~al.} 2010, in American Astronomical Society Meeting Abstracts, Vol. 215, American Astronomical Society Meeting Abstracts \#215, 433.26

\bibitem[{Marscher}(1996)]{marscher_1996}
{Marscher}, A.~P. 1996, in Astronomical Society of the Pacific Conference Series, Vol. 110, Blazar Continuum Variability, ed. H.~R. {Miller}, J.~R. {Webb}, \& J.~C. {Noble}, 248

\bibitem[{Marscher}(2014)]{marscher_2014}
{Marscher}, A.~P. 2014, \apj, 780, 87

\bibitem[{Marscher} {et~al.}(2008)]{marscher_2008}
{Marscher}, A.~P., {Jorstad}, S.~G., {D'Arcangelo}, F.~D., {et~al.} 2008, \nat, 452, 966

\bibitem[{Marscher} {et~al.}(2010)]{marscher_2010}
{Marscher}, A.~P., {Jorstad}, S.~G., {Larionov}, V.~M., {et~al.} 2010, \apjl, 710, L126

\bibitem[Max-Moerbeck {et~al.}(2014)]{moerbeck_2014}
Max-Moerbeck, W., Hovatta, T., Richards, J.~L., {et~al.} 2014, Monthly Notices of the Royal Astronomical Society, 445, 428

\bibitem[{Mead} {et~al.}(1990)]{mead_1990}
{Mead}, A.~R.~G., {Ballard}, K.~R., {Brand}, P.~W.~J.~L., {et~al.} 1990, \aaps, 83, 183

\bibitem[{Otero-Santos} {et~al.}(2023)]{otero_2023}
{Otero-Santos}, J., {Acosta-Pulido}, J.~A., {Becerra Gonz{\'a}lez}, J., {et~al.} 2023, \mnras, 523, 4504

\bibitem[{Pandey} {et~al.}(2022)]{pandey_2021}
{Pandey}, A., {Rajput}, B., \& {Stalin}, C.~S. 2022, \mnras, 510, 1809

\bibitem[{Peterson}(1993)]{peterson_1993}
{Peterson}, B.~M. 1993, \pasp, 105, 247

\bibitem[{Peterson} {et~al.}(1998)]{peterson_1998}
{Peterson}, B.~M., {Wanders}, I., {Horne}, K., {et~al.} 1998, \pasp, 110, 660

\bibitem[Raiteri \& Villata(2021)]{raiteri_2021}
Raiteri, C.~M., \& Villata, M. 2021, Galaxies, 9

\bibitem[{Raiteri, C. M.} {et~al.}(2012)]{raiteri_2012}
{Raiteri, C. M.}, {Villata, M.}, {Smith, P. S.}, {et~al.} 2012, A\&A, 545, A48

\bibitem[{Raiteri} {et~al.}(2013)]{raiteri_2013}
{Raiteri}, C.~M., {Villata}, M., {D'Ammando}, F., {et~al.} 2013, \mnras, 436, 1530

\bibitem[{Raiteri} {et~al.}(2017)]{2017Natur.552..374R}
{Raiteri}, C.~M., {Villata}, M., {Acosta-Pulido}, J.~A., {et~al.} 2017, \nat, 552, 374

\bibitem[{Rajput} {et~al.}(2022)]{rajput_2022}
{Rajput}, B., {Pandey}, A., {Stalin}, C.~S., \& {Mathew}, B. 2022, \mnras, 517, 3236

\bibitem[{Ryan} {et~al.}(2018)]{ryan_2018}
{Ryan}, J.~L., {Siemiginowska}, A., {Sobolewska}, M., \& {Grindlay}, J.~E. 2018, in American Astronomical Society Meeting Abstracts, Vol. 231, American Astronomical Society Meeting Abstracts \#231, 205.04

\bibitem[{Schmidt} {et~al.}(1992)]{schmidt_1992}
{Schmidt}, G.~D., {Stockman}, H.~S., \& {Smith}, P.~S. 1992, \apjl, 398, L57

\bibitem[{Sesar} {et~al.}(2007)]{sesar_2007}
{Sesar}, B., {Ivezi{\'c}}, {\v{Z}}., {Lupton}, R.~H., {et~al.} 2007, \aj, 134, 2236

\bibitem[{Sironi} \& {Spitkovsky}(2014)]{sironi_2014}
{Sironi}, L., \& {Spitkovsky}, A. 2014, \apjl, 783, L21

\bibitem[{Smith} {et~al.}(2009)]{smith_2009}
{Smith}, P.~S., {Montiel}, E., {Rightley}, S., {et~al.} 2009, arXiv e-prints, arXiv:0912.3621

\bibitem[{Udalski} {et~al.}(1997)]{udalski_1997}
{Udalski}, A., {Kubiak}, M., \& {Szymanski}, M. 1997, \actaa, 47, 319

\bibitem[Urry \& Padovani(1995)]{urry_1995}
Urry, C.~M., \& Padovani, P. 1995, Publications of the Astronomical Society of the Pacific, 107, 803

\bibitem[Uttley {et~al.}(2003)]{uttley_2003}
Uttley, P., Edelson, R., McHardy, I.~M., Peterson, B.~M., \& Markowitz, A. 2003, The Astrophysical Journal, 584, L53

\bibitem[{Welsh}(1999)]{welsh_1999}
{Welsh}, W.~F. 1999, \pasp, 111, 1347

\bibitem[{Yang} {et~al.}(2021)]{yang_2021}
{Yang}, S., {Yan}, D., {Zhang}, P., {Dai}, B., \& {Zhang}, L. 2021, \apj, 907, 105

\bibitem[{Zhang} {et~al.}(2022)]{zhang_2022}
{Zhang}, H., {Yan}, D., \& {Zhang}, L. 2022, \apj, 930, 157

\bibitem[Zhang {et~al.}(2023)]{zhang_2023}
Zhang, H., Yan, D., \& Zhang, L. 2023, The Astrophysical Journal, 944, 103

\bibitem[{Zu} {et~al.}(2013)]{zu_2013}
{Zu}, Y., {Kochanek}, C.~S., {Koz{\l}owski}, S., \& {Udalski}, A. 2013, \apj, 765, 106

\bibitem[{Zu} {et~al.}(2011)]{zu_2011}
{Zu}, Y., {Kochanek}, C.~S., \& {Peterson}, B.~M. 2011, \apj, 735, 80

\end{thebibliography}

\label{lastpage}
\end{document}